\documentclass[unsortedaddress,superscriptaddress,10pt,twocolumn,longbibliography]{revtex4-1}
\usepackage[pdftex]{graphicx}
\usepackage{graphicx}
\usepackage[latin1]{inputenc}
\usepackage{amsmath}
\usepackage{ulem}
\usepackage{bbold}
\usepackage[T1]{fontenc}
\usepackage{color,bm, braket}

\usepackage{upgreek}

\newcommand {\I}{\mathrm{i}}
\newcommand {\e}{\mathrm{e}}

\newcommand {\mo}{MoS$_2$}

\newcommand {\pwsum}{\sum\limits}
\newcommand {\pwprod}{\sum\limits}
\newcommand {\pwint}{\int\limits}
\newcommand {\dd}{\mathrm{d}}
\newcommand {\ds}{\displaystyle}
\newcommand {\mathperiod}{\,.}
\newcommand {\mathcomma}{\,,}
\newcommand {\newPsi}{\tilde{\Psi}}

\begin{document}
\DeclareGraphicsExtensions{.pdf}

\title{Tomonaga-Luttinger liquid in a box: electrons confined within \mo~mirror twin boundaries}

\author{Wouter Jolie} 
\email{wjolie@ph2.uni-koeln.de}
\affiliation{II. Physikalisches Institut, Universit\"{a}t zu K\"{o}ln, Z\"{u}lpicher Stra\ss e 77, 50937 K\"{o}ln, Germany}
\affiliation{Institut f{\"ur} Materialphysik, Westf\"{a}lische Wilhelms-Universit\"{a}t M\"{u}nster, Wilhelm-Klemm-Stra{\ss}e 10, 48149 M\"{u}nster, Germany}
\author{Clifford Murray}
\affiliation{II. Physikalisches Institut, Universit\"{a}t zu K\"{o}ln, Z\"{u}lpicher Stra\ss e 77, 50937 K\"{o}ln, Germany}
\author{Philipp S. Wei\ss}
\affiliation{Institut f{\"ur} Theoretische Physik, Universit\"{a}t zu K\"{o}ln, Z\"{u}lpicher Stra\ss e 77, 50937 K\"{o}ln, Germany}
\author{Joshua Hall}
\affiliation{II. Physikalisches Institut, Universit\"{a}t zu K\"{o}ln, Z\"{u}lpicher Stra\ss e 77, 50937 K\"{o}ln, Germany}
\author{Fabian Portner}
\affiliation{Institut f{\"ur} Theoretische Physik, Universit\"{a}t zu K\"{o}ln, Z\"{u}lpicher Stra\ss e 77, 50937 K\"{o}ln, Germany}
\author{Nicolae Atodiresei}
\affiliation{Peter Gr{\"u}nberg Institute and Institute for Advanced Simulation, Forschungszentrum J{\"u}lich, Germany}
\author{Arkady V. Krasheninnikov}
\affiliation{Institute of Ion Beam Physics and Materials Research, Helmholtz-Zentrum Dresden-Rossendorf, 01328 Dresden, Germany}
\affiliation{Department of Applied Physics, Aalto University School of Science, PO Box 11100, FI-00076 Aalto, Finland}
\author{Carsten Busse}
\affiliation{II. Physikalisches Institut, Universit\"{a}t zu K\"{o}ln, Z\"{u}lpicher Stra\ss e 77, 50937 K\"{o}ln, Germany}
\affiliation{Institut f{\"ur} Materialphysik, Westf\"{a}lische Wilhelms-Universit\"{a}t M\"{u}nster, Wilhelm-Klemm-Stra{\ss}e 10, 48149 M\"{u}nster, Germany}
\affiliation{Department Physik, Universit\"{a}t Siegen, 57068 Siegen, Germany}
\author{Hannu-Pekka Komsa}
\affiliation{Department of Applied Physics, Aalto University School of Science, PO Box 11100, FI-00076 Aalto, Finland}
\author{Achim Rosch}
\affiliation{Institut f{\"ur} Theoretische Physik, Universit\"{a}t zu K\"{o}ln, Z\"{u}lpicher Stra\ss e 77, 50937 K\"{o}ln, Germany}
\author{Thomas Michely}
\affiliation{II. Physikalisches Institut, Universit\"{a}t zu K\"{o}ln, Z\"{u}lpicher Stra\ss e 77, 50937 K\"{o}ln, Germany}



\begin{abstract}
Two- or three-dimensional metals are usually well described by weakly interacting, fermionic quasiparticles. This concept breaks down in one dimension due to strong Coulomb interactions. There, low-energy electronic excitations are expected to be bosonic collective modes, which fractionalize into independent spin and charge density waves. Experimental research on one-dimensional metals is still hampered by their difficult realization, their limited accessibility to measurements, and by competing or obscuring effects such as Peierls distortions or zero bias anomalies. Here we overcome these difficulties by constructing a well-isolated, one-dimensional metal of finite length present in \mo\, mirror twin boundaries. Using scanning tunneling spectroscopy we measure the single-particle density of the interacting electron system as a function of energy and position in the 1D box. Comparison to  theoretical modeling provides unambiguous evidence that we are observing spin-charge separation in real space.
\end{abstract}


\maketitle

\section{Introduction}

While long thought to remain a theorist's dream~\cite{Tomonaga1950,Luttinger1963}, a few realizations of one-dimensional (1D) metals suitable for the investigation of low energy excitations as described by the Tomonaga-Luttinger liquid (TLL) theory~\cite{Tomonaga1950,Luttinger1963,Haldane1981} are now available. Among them are metallic carbon nanotubes \cite{Bockrath1999,Ishii2003}, GaAs/AlGaAs based wire devices~\cite{Auslaender2005,Jompol2009,Hashisaka2017}, quasi 1D bulk materials \cite{Claessen2002,Hager2005} and self-assembled atomic wires on semiconductor surfaces~\cite{Segovia1999,Snijders2010,Blumenstein2011,Nakatsuji2012,Ohtsubo2015,Do2015,DeJong2016,Dudy2017}. 

According to TLL theory~\cite{Tomonaga1950,Luttinger1963,Haldane1981}, fingerprints of TLL behavior in 1D metals are power laws for the suppression of the density of states near the Fermi energy $E_F$~\cite{Bockrath1999,Ishii2003,Claessen2002,Hager2005,Segovia1999,Blumenstein2011,Ohtsubo2015} and -- most significantly -- the different dispersions of spin and charge excitations with velocities $v_s$ and $v_c$~\cite{Auslaender2005,Jompol2009,Hashisaka2017,Ma2017}. Their experimental detection is primarily conducted by transport and tunneling transport measurements \cite{Bockrath1999,Auslaender2005,Jompol2009,Hashisaka2017}, angle resolved photoemission electron spectroscopy (ARPES) \cite{Ishii2003,Claessen2002,Segovia1999,Blumenstein2011,Ohtsubo2015,Ma2017}, and scanning tunneling spectroscopy (STS) \cite{Hager2005,Blumenstein2011,Ma2017}.

The difficulties in pinpointing TLL behavior, specifically in self-assembled systems, become apparent by considering the case of self-organized Au wires on Ge(001): From the 1D appearance of the Au adatom chains and power law scaling of the density of states observed by STM and ARPES, TLL behavior was concluded \cite{Blumenstein2011}. In subsequent work \cite{Nakatsuji2012,DeJong2016,Dudy2017} TLL behavior was questioned and even excluded, e.g the 1D character of the system was rejected \cite{Nakatsuji2012,DeJong2016} and the suppression of the density of states was linked to disorder \cite{DeJong2016}. 

These remarks make plain that, in order to gain high-quality data enabling advancement of theory, well-defined 1D systems and new tools to identify TLL behavior are highly desirable. One such new tool is the use of quantum simulators to emulate and explore TLL behavior~\cite{Anthore2018}. As shown below, our approach is the design of an extremely well-defined 1D system of finite length giving rise to a discrete excitation spectrum, accessible by STS.

\begin{figure*}[t!]
	\centering
		\includegraphics[width=0.9\textwidth]{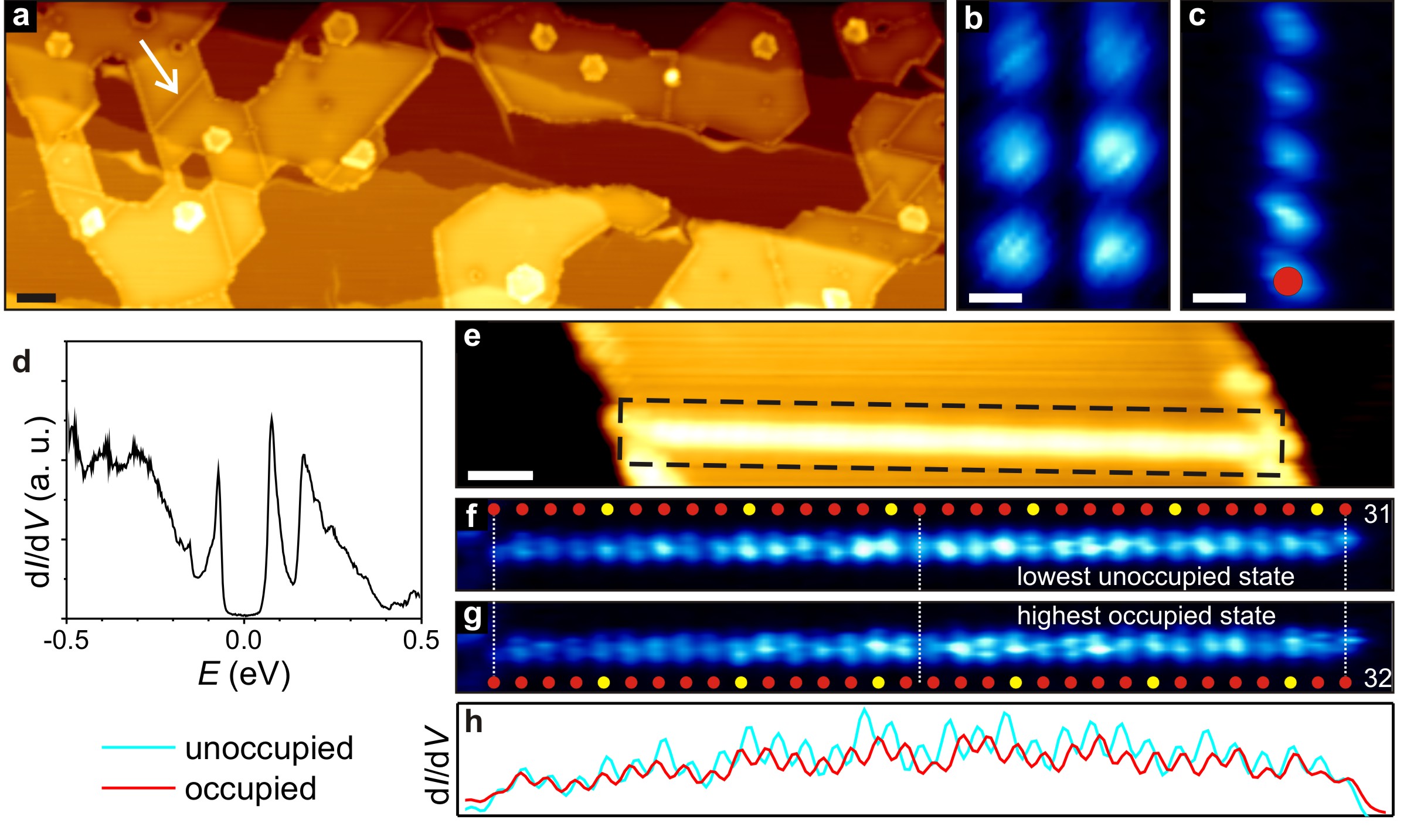}
	\caption{1D states in \mo~mirror twin boundaries. 
	\textbf{a}~Constant current STM topograph ($U=0.9$~V, $I=0.03$~nA, scale bar 10~nm) of a partial \mo~monolayer grown by reactive molecular beam epitaxy on graphene. The latter is fully covering and conformal to the Ir(111) substrate. The \mo~layer islands extend over	several substrate steps and carry small hexagonally-shaped second layer \mo~islands. In consequence of island coalescence and reshaping during the annealing step of the synthesis, straight MTBs are formed along the three dense-packed directions of \mo. They are visible as bright lines of which one is highlighted by a white arrow. 
	\textbf{b}~Constant height STS map of a double line 4{\textbar}4P MTB ($U=0.1$~V, scale bar 0.5~nm). 
	\textbf{c}~Constant height STS map of a single line MTB ($U=0.1$~V, scale bar 0.5~nm). 
	\textbf{d}~d$I$/d$V$ spectrum measured at the red dot position in \textbf{c} ($U_{\rm{stab}}=0.5$~V, $I_{\rm{stab}}=0.1$~nA).  
	\textbf{e}~STM topograph of a single line MTB (U = 0.5~V, I = 0.1~nA, scale bar 2~nm). 
	\textbf{f}~Constant height STS map of the	dashed box area in \textbf{e} at the energy of the lowest unoccupied state 	($U=0.033$~V). 
	\textbf{g}~Constant height STS map of the dashed box area in \textbf{e} at the energy of the highest occupied state ($U=-0.048$~V). 
	The dots marking the maxima in the LDOS patterns make plain that there is one maximum less in the lowest unoccupied state of the MTB.
	From the thin vertical lines in \textbf{f} and \textbf{g} it becomes apparent that the LDOS maxima are in-phase at the edges of the 1D
	box, but in anti-phase in the center.
	\textbf{h}~Line profiles along the MTB, showing the entire phase-relation between lowest occupied and highest unoccupied state.}
	\label{1d_overview}
\end{figure*}

In recent years an exciting realization of a 1D metal was discovered, namely mirror twin boundaries (MTBs) in semiconducting transition metal dichalcogenide monolayers. Their structures were unambiguously determined by transmission electron microscopy \cite{Zhou2013,Lin2015,Wang2016c}, and density functional theory invariably predicted the MTBs to host one-dimensional, metallic states \cite{Zou2013,Le2013,Gibertini2015,Farmanbar2016} that are protected through the large band gap of approximately $2$\,eV in the surrounding 2D-layer. Intense research yielded partly conflicting results regarding the electronic structure of a specific MTB in a monolayer of MoSe$_2$ resting on a van der Waals substrate \cite{Liu2014g,Barja2016,Ma2017}, namely the 4|4P MTB consisting of 4-fold rings sharing a point at the chalcogene site \cite{Zou2013,Komsa2017}. By using room temperature as well as low-temperature (4K) STM and STS, Liu et al. \cite{Liu2014g} found a quantum well state emerging from the finite length of the interpenetrating MTBs. Barja et al.~\cite{Barja2016} proposed a Peierls type charge density wave (CDW) at 4.5~K, but disregarded quantization effects in their MTBs of finite length. By avoiding the CDW regime through room temperature ARPES measurements, Ma et al. \cite{Ma2017} found indications of TLL behavior by observing a suppression of the density of states near $E_F$ and by successfully fitting their spectrum to a Hubbard model with long-ranged interactions.

We go beyond this work and focus on a structurally different MTB in a monolayer of \mo. For this MTB no CDW transition occurs thus allowing us to observe TLL physics down to lowest temperatures. By making 1D wires well isolated from the environment, of high perfection and well-defined length, we are able to observe spin-charge separation in real space through the unique local spectroscopic capabilities of low-temperature STM and STS. This technique can directly probe the probability distribution and energy of discrete TLL excitations in a 1D box. The interpretation of our data is based on the work of Fabrizio and Gogolin \cite{Fabrizio1995} as well as Anfuso and Eggert \cite{Anfuso2003}, who demonstrated that for a TLL in a box the local distribution of the single-particle spectral weight -- determining the probability to inject or extract an electron in a tunneling experiment --  visualizes its fundamental properties.

\section{Results}
\subsection{Design of a 1D box in \mo\,and quantization effects}

To construct our 1D box we grow \mo~islands epitaxially on the van der Waals substrate graphene on Ir(111)~\cite{Hall2017}, see Appendix A. The lower C$_3$ symmetry of \mo~compared to the C$_6$ symmetry of the substrate leads to two equivalent mirror orientations of \mo~islands despite epitaxial alignment. These islands coalescence and reshape during synthesis, resulting in straight MTBs. A white arrow highlights such a MTB in the STM topograph shown in Fig.\,\ref{1d_overview}a. It appears higher than its surrounding when the bias voltage is set close to or within the band gap of \mo~\cite{Hall2017}, consistent with an electronic structure markedly different from the \mo~layer. All MTBs have well-defined lengths, as they terminate at the island edges. Atomic resolution topographs of MTBs are provided in Appendix A.

Two types of MTBs are found in our experiments. As shown in Fig.~\ref{1d_overview}b, one type displays two parallel lines (double line) of dots in an empty state STS map, as has been observed in previous work on MTBs in MoSe$_2$. This MTB has been identified as a 4|4P MTB \cite{Liu2014g,Barja2016,Ma2017}. Depending on the preparation conditions, the 4|4P MTBs make up 5\%-30\% of all MTBs. Their frequency of occurrence reduces with increasing growth temperature, and they are often pinned to defects. This indicates that they are energetically less favorable than the second type of MTB which is predominant in our \mo~samples. This second type of MTB displays in an STS map a single line of dots as visualized in Fig.~\ref{1d_overview}c. The dot periodicity along the line scatters from MTB to MTB, but is close to $3 a$ for the double line and close to $2 a$ for the single line MTB, where $a = 3.15$~\AA~is the lattice parameter of \mo. In the present manuscript we focus on the single line MTB, while the double line MTB is discussed in Appendix A and B, where it is shown to be in fact of 4|4P structure. 

The local density of states (LDOS) $A(E,x)$ present along MTBs is directly accessible with STS, since d$I$/d$V\propto A(E,x)$. Fig.\,\ref{1d_overview}d shows d$I$/d$V$ as a function of the bias voltage $V$ (converted to an energy $E=eV$), measured at the position marked by a red dot in Fig.\,\ref{1d_overview}c. The spectrum reveals a finite density of states throughout the measured energy range, except for a narrow gap $E_{\rm{gap}}$\,of the order of 100~meV located at $E_F$ ($E=0$). All states visible in the spectrum lie within the 2.5~eV band gap of the surrouding \mo~\cite{Hall2017} and hence are strongly confined within the MTB.

Fig.~\ref{1d_overview}e displays an STM topograph of a single line MTB of 20~nm length with terminations formed by the \mo~island edges, whereas Fig.~\ref{1d_overview}f and Fig.~\ref{1d_overview}g are corresponding constant height STS maps of the dashed box in Fig.~\ref{1d_overview}e. Both maps are measured at the peak energies of lowest unoccupied and highest occupied state, respectively. As follows from the careful comparison of the two patterns, the number of maxima increases by one when moving from the lowest unoccupied state to the highest occupied state. The uniform spacing of the maxima and their in-phase relation at the box edges imply an anti-phase relation in the middle of the MTB, as seen best in the corresponding linescans shown in Fig.~\ref{1d_overview}h. 

This pattern is exactly what is expected for non-interacting particles (holes) in a box of size $L$: the wave number $k$ of the highest occupied and lowest unoccupied state differ by $\pi/L$, leading to an extra maximum in the resulting probability distribution $A(E,x)\sim \sin^2(k x)$. In the following, we will first show that DFT calculations reproduce both the shape and phase relation of the bound-state wave functions near $E_F$. Then we will discuss the role of interaction effects which are mandatory to accurately reproduce our findings.

\begin{figure*}[t!]
	\centering
		\includegraphics[width=1\textwidth]{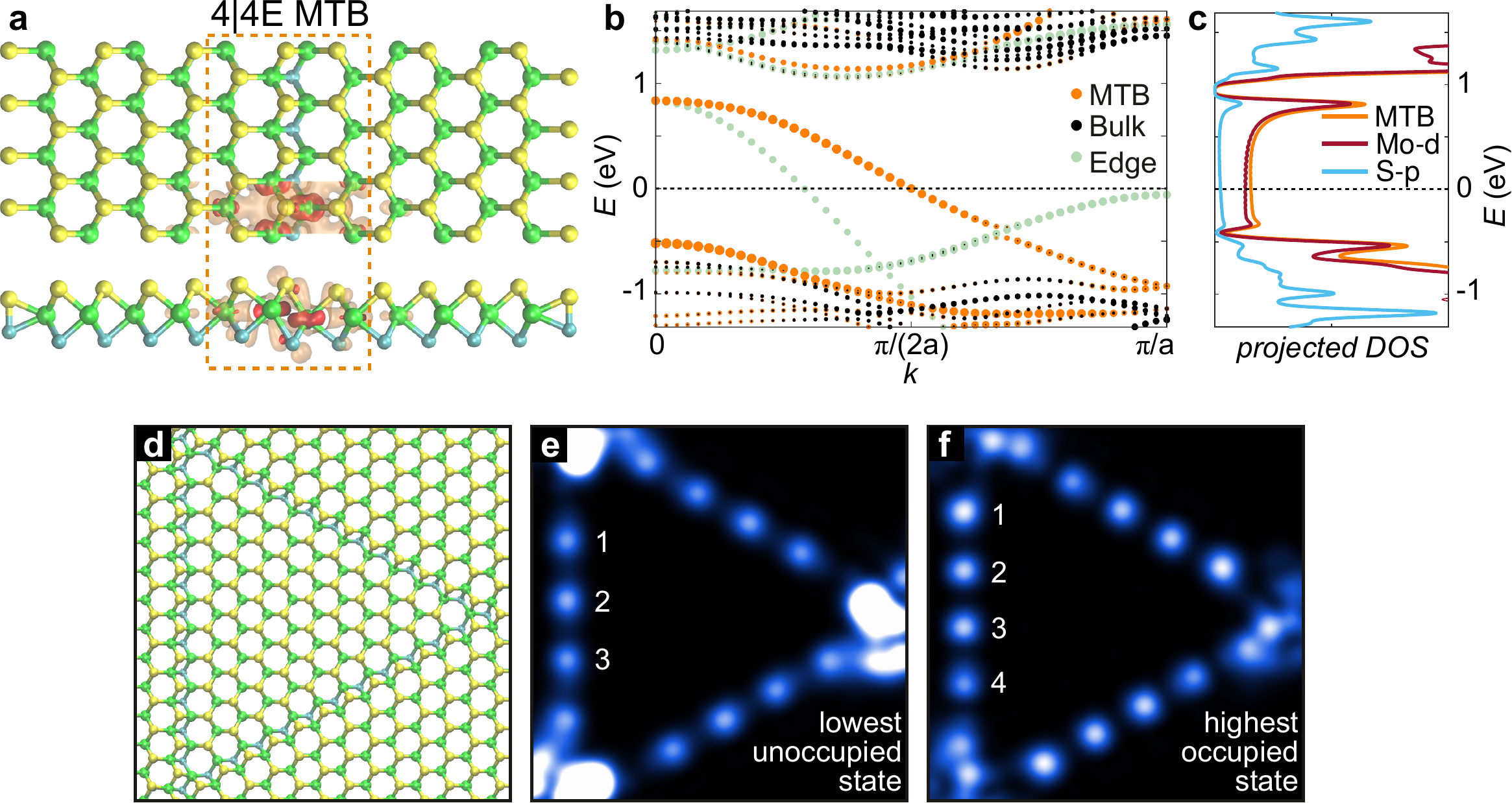}
\caption{DFT calculations for 4{\textbar}4E mirror twin boundaries in \mo. 
	\textbf{a} Top and side view of a ball-and-stick model. Mo atoms: green. S atoms: yellow (top layer) and cyan (bottom layer).  
	\textbf{b} Band structure calculated for the ribbon geometry of \textbf{a} with periodic boundary conditions in the direction along the MTB. Horizontal dashed line denotes the position of $E_F$ at $E = 0$. The hole-like band present at the 4{\textbar}4E MTB is colored orange and crosses	$E_F$ at $k \approx \frac{\pi}{2a}$. The partial charge density of the MTB band wave functions at $k=\pi/(2a)$ is shown in \textbf{a} with two different isosurface values in red and light red. Bulk bands are colored black and bands located at the ribbon edges specific to the finite sized supercell are colored green.
	\textbf{c} Projected density of states (orange) of the atoms around the MTB within the red dashed box in \textbf{a} corresponding to the metallic MTB band in \textbf{b}, and its Mo-$d$ (dark red) and S-$p$ (light blue) contributions.
	\textbf{d} Geometry of triangular inversion domain enclosed by three 4{\textbar}4E MTB segments. 
	\textbf{e,f} LDOS maps (simulated STS maps) for	the supercell shown in \textbf{d} at a height of 2.8\,\AA. Discrete states with wave vector just below (above) $k_F$ corresponding to lowest unoccupied (highest occupied) state are shown in \textbf{e} (\textbf{f}). $E_F$ was adjusted to match $k_F = \frac{\pi}{2a}$. From comparison of \textbf{e} and \textbf{f} it is apparent that the number of maxima on a 4{\textbar}4E MTB segment increases by one when moving from the lowest unoccupied state to the highest occupied state. Note that the features in the corners of the LDOS maps in \textbf{e} and \textbf{f} are due to the interactions of the triangle corners in the DFT supercell.}
	\label{1d_dft}
\end{figure*}

\subsection{Density functional theory calculations}

We propose the single line MTB to possess the 4|4E structure, i.e. to consist of 4-fold rings which share an edge, 
as schematically depicted in the ball and stick model of Fig.~\ref{1d_dft}a. 
The DFT calculated band structure for the ribbon geometry of Fig.~\ref{1d_dft}a is displayed in Fig.~\ref{1d_dft}b. 
Apparent is a hole-like band localized at the 4|4E MTB, with its maximum at $k = 0$ and crossing $E_F$ at $k = k_F \approx \frac{\pi}{2a}$. 
The wave functions related to this band at $k=\frac{\pi}{2a}$ are illustrated by the partial charge density isosurface plots in Fig.~\ref{1d_dft}a,
and show clear localization to Mo- and S-atoms around the MTB. Furthermore, the angular momentum projections evidence
that they have purely Mo-$d$ and S-$p$ character, as illustrated in the projected DOS in Fig.~\ref{1d_dft}c.
From the partial charge density plots in Fig.~\ref{1d_dft}a it is also obvious that
the tunneling current stems primarily form the S-$p$ orbitals localized at the S atoms, as these orbitals extend furthest into the vacuum. We note that DFT calculations including spin-orbit coupling show the metallic MTB band unchanged and spin-degenerate.
For an inversion domain supercell bounded by three 4|4E MTB segments as shown in Fig.~\ref{1d_dft}d, the simulated LDOS maps for the lowest unoccupied (Fig.~\ref{1d_dft}e) and highest occupied (Fig.~\ref{1d_dft}f) states matches in appearance and periodicity precisely with the d$I$/d$V$ maps in Fig.~\ref{1d_overview}c,f,g. Noteworthy, also the number of LDOS maxima increases by one upon moving from the lowest unoccupied to the highest occupied state. Our assignment of the single line being a 4|4E MTB is backed up by the fact that DFT calculations also reproduce STS maps of the 4|4P double line MTB as documented in Appendix~B.

\begin{figure*}[t!]
	\centering
		\includegraphics[width=0.9\textwidth]{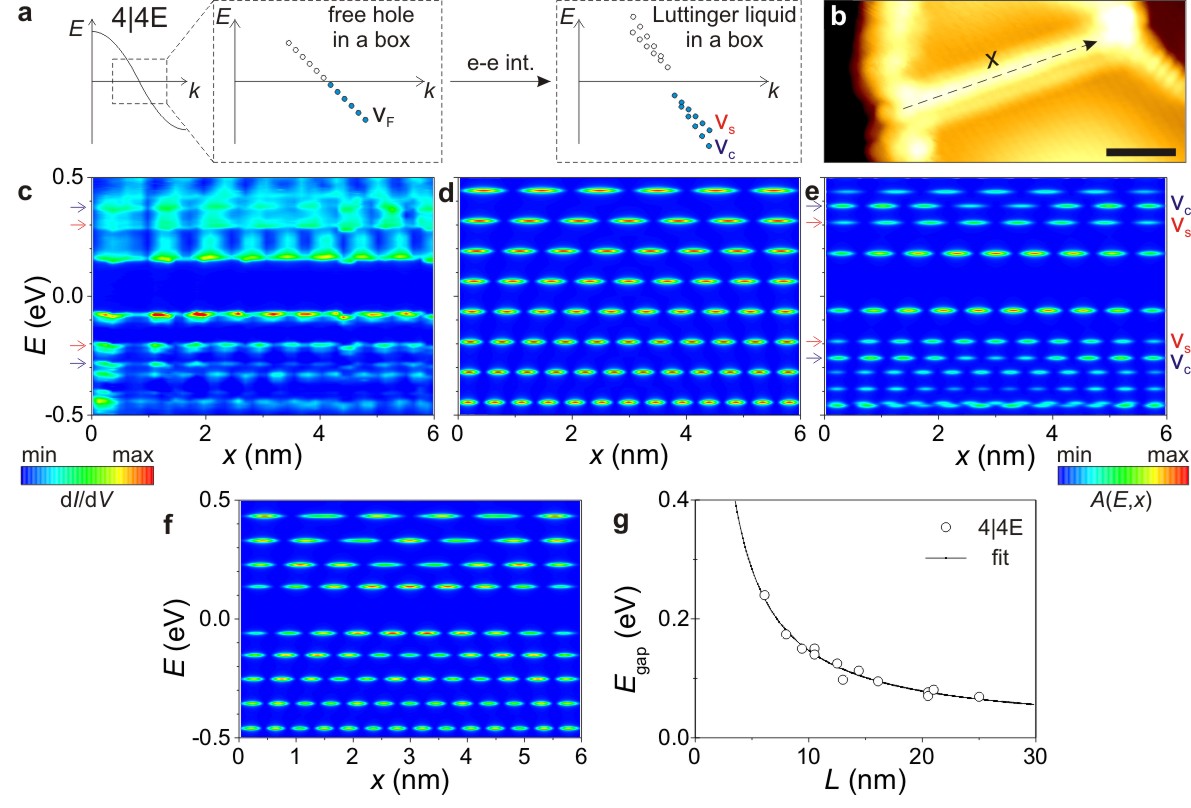}
	\caption{Confined quasiparticles in \mo~MTBs. 
	\textbf{a} Sketch of the TLL model: The hole-like band of the 4{\textbar}4E MTB is first linearized, then quantized. Filled states are occupied. Including electron-electron interactions increases the gap at the chemical potential and lifts the degeneracy of the charge and spin quasiparticles. 
	\textbf{b} STM topograph of a 4{\textbar}4E MTB ($U=1$~V, $I=0.2$~nA, scale bar 2~nm). Dotted line shows the path of the tip during STS data acquisition. 
  \textbf{c} Corresponding d$I$/d$V$ signal as a function of energy and position ($U_{\rm{stab}}=0.8$~V, 
$I_{\rm{stab}}=1$~nA). Arrows denote states which have the same number of maxima and are in-phase throughout the MTB. Color scale ranges from dark blue for zero d$I$/d$V$ signal to red for maximum d$I$/d$V$. Same color scale is used in \textbf{d} to \textbf{f} for the simulated LDOS.
	\textbf{d} Simulated LDOS assuming non-interacting holes confined to a 1D box using the band structure from DFT. 
	\textbf{e} Simulated LDOS using the TLL model (${v_c = 0.38~\mathrm{nm \cdot eV}}$, ${v_s = 0.25~\mathrm{nm\cdot eV}}$, $K_c =  0.5$). First spin and charge excitations with velocities $v_s$ and $v_c$ are highlighted by arrows. They display identical number of maxima and in-phase behavior over the entire MTB length, as observed experimentally. 
\textbf{f} Simulated LDOS assuming a CDW with an interaction strength matched to fit the experimentally observed gap. 
\textbf{g} Dependence of $E_{\rm{gap}}$ as a function of 4{\textbar}4E MTB length $L$, supporting our TLL interpretation.
}
	\label{realspace}
\end{figure*}

\subsection{Failure of the free particle in a 1D box picture}

While our DFT calculations seem to suggest a free particle (hole) in a 1D box picture, as schematically sketched to the left of Fig.~\ref{realspace}a, this model fails in reproducing the quantized electronic structure in our spectra. A sequence of STS spectra along the 4|4E MTB (measurement path is indicated by a dotted line) shown in Fig.~\ref{realspace}b reveals in the color plot of Fig.\,\ref{realspace}c the presence of additional, well-separated quantized states above the lowest unoccupied state and below the highest occupied state. 

In Fig.\,\ref{realspace}d we show the corresponding pattern expected for non-interacting holes confined to a 1D box of the same length using the band-structure from DFT. Several discrepancies are present compared to Fig.\,\ref{realspace}c: In the experiment (i) the energy gap $E_{\rm{gap}}$ between the highest occupied and the lowest unoccupied state is much larger compared to the neighboring energy level spacings; (ii) there is a higher number of energy levels and the level spacing is not approximately equidistant; (iii) most striking, some states adjacent in energy display the same number of maxima and hence are in-phase throughout the entire MTB (compare states highlighted by arrows in Fig.\,\ref{realspace}c). In the following, we will argue that these observations can be explained by the TLL theory.

\subsection{Tomonaga-Luttinger liquid in a 1D box}

TLL theory is an effective field theory describing the low-energy excitations of an interacting 1D metal. The low-energy excitations are not Fermi liquid quasiparticles but instead spin- and charge density waves which travel with two different velocities, $v_s$ and $v_c$, giving rise to two different dispersions, as schematically sketched in Fig.\,\ref{realspace}a. When an electron is injected into the system, it fractionalizes: it creates a multitude of spin- and charge excitations. In a finite size system these excitations are standing waves with discrete energies which provide characteristic fingerprints in an STM experiment. 

At low energies, the Hamiltonian of the Luttinger liquid in the finite system of length $L$ can be written as
\begin{eqnarray} \label{eq:Hamiltonian_bosons}
H 
	&=& \frac{(N-N_0)^2}{2 c L}+\frac{\pi v_s S_z^2}{K_s L} \nonumber \\
	&&+
		\pwsum_{m>0} 
		\left(
		v_s q_m b^\dagger_{s,m} b_{s,m}+v_c q_m b^\dagger_{c,m} b_{c,m}
		\right)
\end{eqnarray}
where $b^\dagger_{s,m}$ and $b^\dagger_{c,m}$ are the creation operators of a spin- and charge excitation with quantum number $m \in \mathbb{N}$ and $c$ the capacity per length of the wire.
Due to the finite size of the system, the excitations have discrete energies ${v_s q_m}$ and ${v_c q_m}$, where  ${q_m=\frac{\pi}{L} m}$,  is the discrete momentum which is defined to be positive. The relation of the bosonic field to the fermionic operators depends on two Luttinger parameters, $K_s$ and $K_c$, which encode the effects of interactions. Note that in the finite size system, positive and negative momenta (and left- and right movers) are always coupled by the boundary conditions. 

An important ingredient of the Luttinger liquid theory are the first two terms in Eq.~\eqref{eq:Hamiltonian_bosons}, which describe the so-called zero-modes. The first term is simply the charging energy, $N$ is the total charge in the box, $N_0$ is a background charge parametrizing the chemical potential. This Coulomb-blockade barrier determines $E_{\rm{gap}}$. It contains contributions from the finite size level spacing, the local interaction, and the long-ranged Coulomb interaction. In plain-vanilla TLL (a theory with purely local interactions) the capacity per length $c$ is fixed by the Luttinger liquid parameter $K_c$, $c=2 K_c/(\pi v_c)$. We consider the capacitance of the grain boundary as an independent fitting parameter arising from the long-ranged Coulomb interaction \cite{Boese2001}. The second term is a similar contribution in the spin sector, where we assume that the ground state has zero spin. In our calculations we set $K_s=1$ (assuming spin-rotation invariance).

In order to compare to experiment we calculate the LDOS of the Luttinger liquid:
\begin{equation} \label{eq:LDOS_GF}
\begin{array}{rcl}
A(E,x)
	& = &
	 \ds\pwsum_{\sigma}
	 \pwint_{-\infty}^{\infty} \! \! \dfrac{\mathrm{d}t}{2 \pi} \,
	 \mathrm{e}^{\mathrm{i} E t}
	 \Braket{ \left\{ \Psi_{\sigma}(x,t), \Psi_{\sigma}^\dagger(x,0)\right\} }
	 
\,,
\end{array}
\end{equation}
where $\Psi_{\sigma}$ are the fermionic field operators.
${\left\{\cdot,\cdot \right\}}$ denotes the anticommutator
and $\Braket{...}$ is the expectation with respect to the $N_0$-fermion ground state, i.e.~a state without bosonic excitations,
see Appendix~C for details.

The spectrum predicted by TLL is shown as color plot in Fig.~\ref{realspace}e for parameters matching the experimental findings in Fig.~\ref{realspace}c. Our Luttinger liquid in a box overcomes all discrepancies mentioned above, yielding (i) the proper $E_{\rm{gap}}$, (ii) the increased number of levels with non-equidistant spacings, and (iii) adjacent states with the same number of maxima and an in-phase relation of the maxima throughout the entire MTB length. These states, highlighted by arrows in Fig.~\ref{realspace}e, result from the different velocities $v_s$ and $v_c$ of the first spin and charge excitations, which are well separated in energy.

\subsection{Absence of a Peierls-type charge density wave}

Our data cannot be explained by a Peierls-type CDW suggested for the 4{\textbar}4P MTB in MoSe$_2$ \cite{Ma2017,Barja2016}. A calculated spectrum for the MTB with a CDW matched to reproduce $E_{\rm{gap}}$ is represented in Fig.~\ref{realspace}f (see Appendix D). It is obvious that the discrepancies (ii) and (iii) remain. Moreover, we find that $E_{\rm{gap}}$ decreases with $1/L$ consistent with TLL theory, where $E_{\rm{gap}}=\left(\frac{1}{c} +\frac{\pi v_s}{2} \right) \frac{1}{L}$, but not with a CDW scenario, where $E_{\rm{gap}}=\rm{const.}$ 
 A fit to $E_{\rm{gap}}(L) = A/L + B$ (Fig.~\ref{realspace}g) leads to a tiny extrapolated gap $E_{\rm{gap}}(\infty)=(10\pm6)$~meV. Assumption of a Peierls-type CDW gap would be inconsistent with our room temperature observation of the beating pattern, see Appendix D. The value of $A=(1.37 \pm 0.07)\,\rm{eV\,nm}$\,is consistent with the estimate $A = \frac{\pi v_s}{2} + \frac{\pi v_c}{2 K_c} = (1.6 \pm 0.3)\,\rm{eV\,nm}$, see Sec. IID. 

\begin{figure*}[t!]
	\centering
		\includegraphics[width=0.7\textwidth]{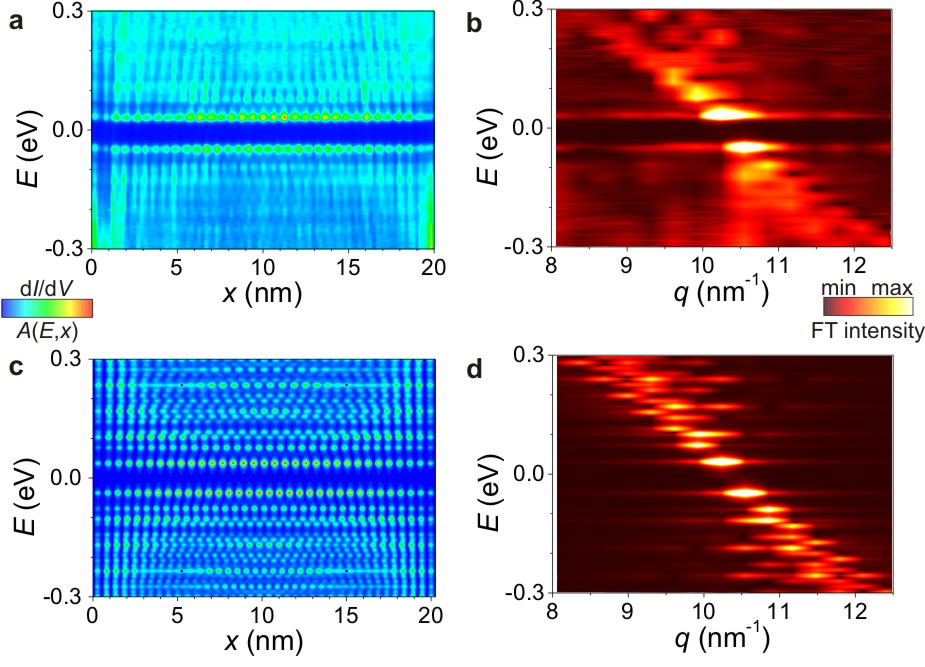}
	\caption{Dispersion of confined quasiparticles in \mo~MTBs. 
	\textbf{a} Color plot of the d$I$/d$V$ signal along the MTB displayed in Fig.~\ref{1d_overview}e as a function of	energy and position ($U_{\rm{stab}}=0.5$~V, $I_{\rm{stab}}=0.1$~nA).
	\textbf{b} FT of the experimental data, revealing the dispersion of the	confined quasiparticles. 
	\textbf{c} Simulated LDOS using our TLL model (${v_c = 0.45~\mathrm{nm \cdot eV}}$, ${v_s = 0.27~\mathrm{nm\cdot eV}}$, $K_c=0.5$). Same color scale as used in \textbf{a}.
	\textbf{d} FT of the simulated LDOS in \textbf{c} with the same color scale as in \textbf{b}.}
	\label{1d_confinement}
\end{figure*}

\subsection{Spin-charge separation in reciprocal space}

Additional insight and support for the TLL picture is provided by the analysis of the d$I$/d$V$ spectra taken along longer 4|4E MTBs. Fig.~\ref{1d_confinement}a and b present a color plot of the MTB of Fig.~\ref{1d_overview}d together with its Fourier transform (FT). The latter is considerably more instructive, since it directly reveals the dispersion of the confined states~\cite{Meyer2003}. Corresponding simulations using TLL theory are shown in Figs.~\ref{1d_confinement}c and d.

The strongest peaks in the simulated FT of Fig.\,\ref{1d_confinement}d are located at the energies $\pm\frac{E_{\rm{gap}}}{2}$ of the highest occupied and lowest unoccupied states. Their corresponding momenta are $2 k_F^\mp$ where $k_F^-$ is the momentum of the highest occupied state and $k_F^+=k_F^- + \frac{\pi}{L}$ is the momentum of the lowest unoccupied state. They describe excitations
where only the fermion number (zero-modes of the TLL) changes and no spin- or charge modes are excited. A series of prominent peaks is located at momenta $2 (k_F^{\pm} \pm  q_m) $, $q_m=\frac{\pi}{L} m$, $m> 0$, with energies $\mp (\frac{E_{\rm{gap}}}{2} + v_s q_{m})$, $\mp (\frac{E_{\rm{gap}}}{2} + v_c q_{m})$ describing a situation where the injected electron produces a pure spin or a pure charge excitation. Fitting these peaks can be used to determine the charge and spin velocities, $v_c$
and $v_s$, assuming that $v_s<v_c$ as expected for repulsive interactions. This situation is depicted in Fig.~\ref{realspace}a. 

It is also possible that spin- and charge excitations are present simultaneously. When several modes with momenta $q_1,q_2,\dots, q_n>0$ are excited, this leads to peaks at the momenta $2 (k_F^\pm  \pm \sum_i \sigma_i q_i)$ with arbitrary signs $\sigma_i=\pm 1$. The associated energies are $\mp \left(\frac{E_{\rm{gap}}}{2} +\sum_i v_{\nu} q_{i_{\nu}}\right)$ with $\nu=c/s$ for a charge/spin excitation, respectively. These extra peaks are, however, only activated when one of the Luttinger liquid parameters $K_c$ and $K_s$ deviates from $1$, where $1$ corresponds to the non-interacting case. In our simulations $K_c = 0.5$ and 
$K_s = 1$. The smaller the Luttinger parameters are, the more weight is transferred from the primary peaks to the side peaks giving rise to a more symmetrical spectrum with an accumulation of weight below and above the two strongest peaks. Our data does not allow a determination of $K_c$ with high precision; we estimate $K_c = 0.5 \pm 0.1$ (see Appendix E for spectra with different values of $K_c$).

Comparing the simulated FT plot of Fig.~\ref{1d_confinement}d to the corresponding experimental FT plot of Fig.~\ref{1d_confinement}b makes plain that in the experimental data, though more blurred, the spectral weight is not only aligned along a simple curve $v_F k$, in contrast to the non-interacting case. Clear indications for the presence of a second velocity are visible. Furthermore, extra weight is accumulated above and below the central peaks, in coincidence with the theoretical modeling.

\section{Discussion and Conclusions}

While we find a good qualitative agreement between experiment and theory, there are also discrepancies. Most prominently, the theory is by construction exactly particle-hole symmetric, while in the experiment both the position of the peaks and their width differ for positive and negative energies. In Fig.~\ref{realspace}c, the peaks at positive energies are strongly broadened. The coupling to electrons in the substrate is the prime candidate for this effect. One further effect has been neglected in our theoretical analysis: backscattering in the spin channel, which cannot be treated exactly within TLL theory. While backscattering is formally irrelevant and vanishes for $L\to \infty$, it is only logarithmically suppressed as function of $L$ for spin-rotation symmetry. It is therefore expected to affect the finite size spectrum, possibly explaining some of the discrepancies in peak positions.

The insight obtained here for the 4|4E MTBs in \mo~sheds also some light on the controversial results for the 4|4P MTB in MoSe$_2$. Quantum confinement effects, as detected by Liu et al. \cite{Liu2014g}, are a necessary consequence of a finite MTB wire length. Thus, for a finite length MTB with a metallic band crossing $E_F$ around $k = \pi/3a$ (as for the 4|4P MTB in MoSe$_2$ \cite{Komsa2017}), the observations of a gap around $E_F$, of an approximate period tripling, and an anti-phase relation between the highest occupied and lowest unoccupied state in the center of the wire are to be expected, and do not constitute evidence for the presence of a CDW as proposed by Barja et al. \cite{Barja2016}. Based on the similarity of the 4|4P and 4|4E MTBs and our clear cut evidence for the presence of a TLL in the 4|4E MTB in \mo, it appears likely that indeed a TLL is also present in the 4|4P MTB in the monolayer of MoSe$_2$ as proposed by Ma et al. \cite{Ma2017}.

When comparing previous results on the 4|4P MTB in MoSe$_2$ to ours on the 4|4E MTB in \mo, it turns out that the Luttinger parameter $K_c$ in the range of 0.20 to 0.21 obtained by Ma et al. \cite{Ma2017} is much lower than our estimate $K_c = 0.5 \pm 0.1$. One possible reason for this discrepancy is that Ma et al. deduce $K_c$ from a power law fit to the density of states near $E_F$ by averaging over a dense network of 1D subsystems. This analysis does not take into account the finite wire length between crossing points (of the order of 10\,nm). The finite wire length implies an extra suppression of the density of states both due to finite size gaps (see our Fig.~\ref{realspace}g) and due to an extra suppression of the density of states close to defects and walls predicted by Luttinger liquid theory \cite{Meden2000}. This extra suppression of the density of states may lead to an estimate of $K_c$ which is systematically too small.

In conclusion, STS spectral maps along MTBs and their FTs show clear evidence for spin-charge separation, characteristic for a quantum confined TLL. We envision that higher resolution data could be obtained by further decoupling the 1D metal from its environment and by lowering the temperature, enabling a quantitative comparison to TLL theory. Moreover, chemical gating and defect engineering of the MTBs might enable one to modify the correlation strength in the TLL or even create new exotic phases.


\begin{acknowledgments}
W. J., C. M., P. S. W., J. H., N. A., A. R., and T. M. gratefully
acknowledge financial support from the Deutsche Forschungsgemeinschaft
(DFG) through the CRC 1238 within projects A01, B06, C01 and C02. W. J.
 acknowledges support from the Bonn-Cologne Graduate School for Physics
and Astronomy. This work is also supported by the Cologne University via the Advanced Postdoc Grant "2D materials beyond graphene". Support from the Academy of Finland is acknowledged under Projects No. 286279 (A. V. K. and H.-P. K.) and 311058 (H.-P. K.). We also thank CSC-IT Center for Science Ltd. and generous grants of computer time and PRACE for awarding us access to Hazel Hen at High Performance Computing Center, University of Stuttgart, Germany. We acknowledge useful discussions with J. Sirker and S. Trebst.
\end{acknowledgments}

\appendix
\section{Experimental methods and high-resolution images of \mo}

\subsection{Sample preparation}
The synthesis of \mo~on the substrate graphene on Ir(111) is conducted in a two-step process~\cite{Hall2017}: During the first step, Mo is evaporated from a rod with a rate of $0.125$~monolayers/min on the graphene surface at room temperature in a S pressure of $p\approx5\times10^{-9}$~mbar.

During the second step, the sample is annealed for 5 minutes at $T=1050$~K in a S pressure of $p\approx5\times10^{-9}$~mbar. This leads to large, flat monolayer \mo~islands with small second layer islands on top.

\subsection{Scanning tunneling microscopy}
All STM and STS experiments were conducted at $T=5~K$. For STS, we measure the d$I$/d$V$~signal using the lock-in technique (modulation voltage $V_{\rm{mod}}=4$~mV; frequency $f=777$~Hz). For our color plots, we use a linear interpolation of the discrete data.

\subsection{Identifying mirror twin boundaries}

\begin{figure}[t!]
	\centering
		\includegraphics[width=0.475\textwidth]{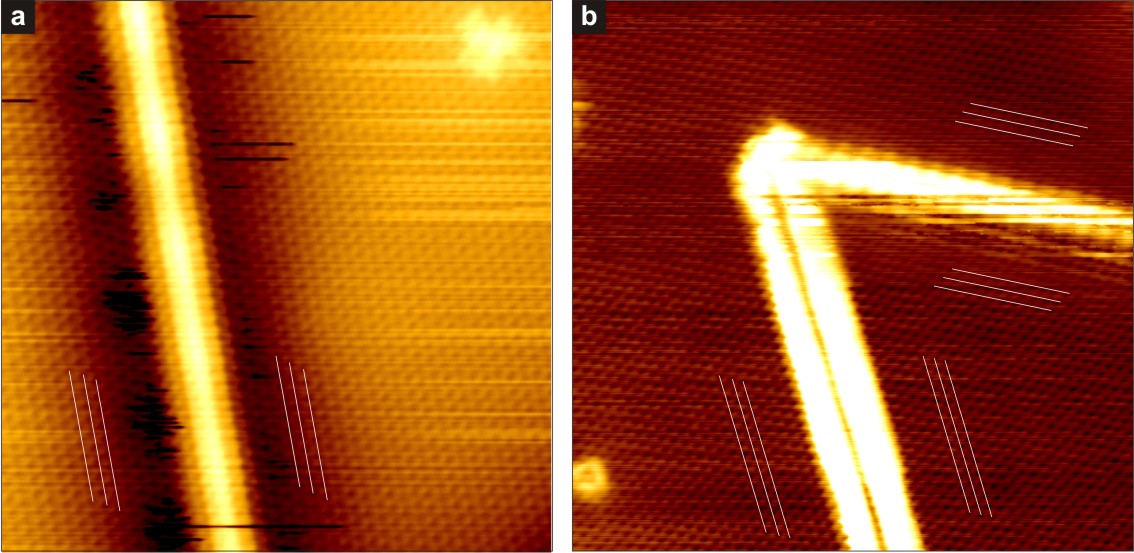}
	\caption{MTBs in monolayer \mo. \textbf{a} Constant current STM image of a 4{\textbar}4E MTB in \mo/graphene/Ir(111) ($U=0.9$~V, $I=0.5$~nA, image size $\left(12\times12\right)~\rm{nm}^2$). The white lines show that both grains are perfectly alined. \textbf{b} Constant current STM image of a grain boundary forming both a 4{\textbar}4E and a 4{\textbar}4P MTB ($U=0.4$~V, $I=1$~nA, image size $\left(15\times15\right)~\rm{nm}^2$).}
	\label{mo_atoms}
\end{figure}

Atomically resolved STM images of MTBs in monolayer \mo/graphene/Ir(111) are shown in Fig.~\ref{mo_atoms}. In Fig.~\ref{mo_atoms}a, one finds two islands separated by a line defect. The densed packed rows in both islands have the same orientation -- visualized with white lines in Fig.~\ref{mo_atoms}a. Hence this line defect must be a MTB. Together with its electronic signatures, which are described in the main text, the MTB can be attributed to a 4{\textbar}4E MTB.

Another atomically resolved STM image is shown in Fig.~\ref{mo_atoms}b. It displays a boundary containing two different line defects which meet at an angle. One has a single line structure as in Fig.~\ref{mo_atoms}a, while the other exhibits a double structure. It is this double line feature, together with a periodic beating of approximately 3a (not visible here), which clearly distinguishes this boundary. The orientation of the atomic rows on both sides of the double line is identical. Hence both boundaries must be MTBs. The MTB with double lines is attributed to a 4{\textbar}4P MTB.


\section{Density functional theory}

\subsection{Density functional theory calculations}
All density functional theory calculations were carried out within the plane-wave basis and the projector augmented wave framework as implemented in VASP~\cite{Kresse1993,Kresse1996}. The exchange-correlation effects are treated with the functional proposed by Perdew, Burke, and Ernzerhof (PBE)~\cite{Perdew1996}. The atomic models in the ribbon calculations consist of tetragonal cells with sizes of about $6 \sqrt(3) \times 1 \times 1$ and periodic only along the MTB with vacuum claddings in the other two directions. The 4{\textbar}4E or 4{\textbar}4P MTB are located symmetrically in the middle, while the ribbon edges are of the S$_2$-passivated Mo-zigzag type. The $k$-point sampling of $1 \times 12 \times 1$ and the plane-wave cutoff of 500 eV guarantee converged total energies. For MTBs of finite length, we adopted models where triangular MTB loops are embedded within a $16 \times 16$ supercell. In this case, $\Gamma$-point sampling and reduced plane-wave cutoff are adopted. The STS images are obtained within the Tersoff-Hamann approximation~\cite{Tersoff1985}.

\begin{figure}[t!]
	\centering
		\includegraphics[width=0.475\textwidth]{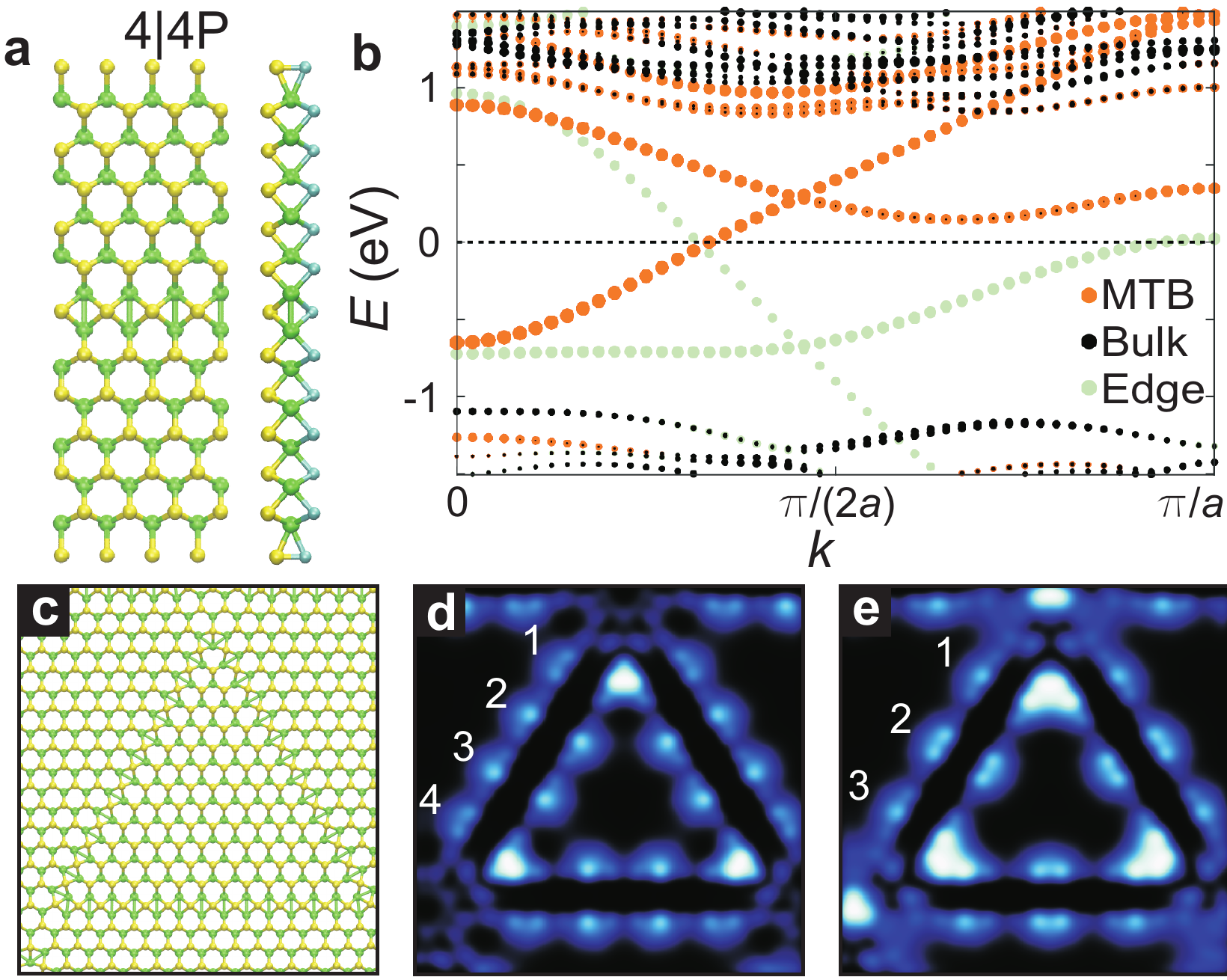}
	\caption{DFT calculations for 4{\textbar}4P mirror twin boundaries in \mo. 
	\textbf{a} Top and side view of a ball-and-stick model. Mo atoms: green. S atoms: yellow (top layer) and cyan (bottom layer). 
	\textbf{b} Band structure calculated for the ribbon geometry of \textbf{a} with periodic boundary conditions in direction along MTB. Horizontal dashed line denotes the position of $E_F$ at $E = 0$. The electron-like band present at the 4{\textbar}4P MTB is colored orange and crosses	$E_F$ at $k \approx \pi/(3a)$. A second band appears above $E_F$. Bulk bands are colored black and bands located at the ribbon edges specific to the finite sized supercell are colored green.	
	\textbf{c} Geometry of triangular inversion domain enclosed by three 4{\textbar}4P MTB segments. 
	\textbf{d,e} LDOS maps (simulated STS maps) for	the supercell shown in \textbf{c} at a height of 2.8\,\AA. Discrete states with wave vector just above (below) $k_F$ corresponding to lowest unoccupied (highest	occupied) state are shown in \textbf{d} (\textbf{e}). $E_F$ was adjusted to match $k_F = \pi/(3a)$.}
	\label{44P_DFT}
\end{figure}

\subsection{4{\textbar}4P mirror twin boundaries}

We propose that the double line MTB possesses the 4{\textbar}4P structure, i.e. consists of 4-fold rings which share a point, as schematically depicted in the ball and stick model of Fig.~\ref{44P_DFT}a. The DFT calculated band structure for the ribbon geometry of Fig.~\ref{44P_DFT}a is displayed in Fig.~\ref{44P_DFT}b. Apparent is an electron-like band localized at the 4{\textbar}4P MTB, with its minimum at $k = 0$ and crossing $E_F$ at $k = k_F \approx \pi/(3a)$. For an inversion domain supercell bounded by three 4{\textbar}4P MTB segments as shown in Fig.~\ref{44P_DFT}c, the simulated LDOS maps for the lowest unoccupied (Fig.~\ref{44P_DFT}d) and highest occupied (Fig.~\ref{44P_DFT}e) state match in appearance and periodicity precisely with the STS maps of the 4{\textbar}4P MTBs. Also to be noted, is that the number of LDOS maxima decreases by one upon moving from the lowest unoccupied to the highest occupied state. Our assignment of the double line being a 4{\textbar}4P MTB is in agreement with literature~\cite{Liu2014g,Barja2016,Ma2017}. 


\section{Tomonaga Luttinger liquid theory in a 1D box}

To describe a Luttinger liquid in a system of finite length \cite{Fabrizio1995,Anfuso2003}, the electron operator is first split into a left-moving and a right-moving part:

\begin{equation}
\Psi_{\sigma}(x)=e^{-i k_F x} \Psi_{\sigma,L}(x)+e^{i k_F x} \Psi_{\sigma,R}(x).
\end{equation}

 At the two boundaries, a right-moving electron is scattered into a left-moving one.
It is therefore  useful to define a new field $\newPsi_{\sigma}(x)$ which is $\Psi_{\sigma,R}(x)$ for $0\le x \le L$ and $-\Psi_{\sigma,L}(-x)$ for $-L\le x \le 0$. This new field has simple periodic boundary conditions $\newPsi_{\sigma}(-L)=\newPsi_{\sigma}(L)$ and naturally incorporates the physics at the boundary.

The electron operator is obtained from the bosonization identity:
\begin{eqnarray} \label{eq:bosonization-id}
\newPsi_{\sigma}(x,t) 
&=& 
\dfrac{F_{\sigma}(x,t)}{\sqrt{2 \pi a}}
\e^{\I \Phi_c (x,t)}
\e^{\I \sigma \Phi_s (x,t)}
\mathperiod \\
\Phi_{\nu}(x,t) 
&=& 
\pwsum_{m=1}^{\infty}
\dfrac{1}{\sqrt{m}}
\left(
\chi_{\nu,m}(x) \e^{-\I v_{\nu} q_{m} t} b_{\nu,m} + \mathrm{h.c.}
\right)
\mathcomma \\
F_{\sigma}(x,t)
&=&
F_{\sigma}
\e^{\I \frac{E_{\rm gap}}{2} t}
\e^{\I (\frac{\pi}{L}x - E_{\rm gap} t) (N_{\sigma} - N_{0,\sigma}) }
\mathcomma
\end{eqnarray}
with Klein factors $F_\sigma$ and
${\chi_{\nu,k}(x) 
= (
\alpha_{\nu}\e^{\I q_{m_{\nu}} x}
-\beta_{\nu}\e^{-\I q_{m_{\nu}} x}
)/\sqrt{2}}$ 
and 
${\alpha_{\nu} 
= 
( K_{\nu}^{1/2} + K_{\nu}^{-1/2} )/2 }$,
${
\beta_{\nu} 
= 
( K_{\nu}^{1/2} - K_{\nu}^{-1/2})/2 }$ (${\nu = c,s}$).

In order to evaluate the LDOS \eqref{eq:LDOS_GF}
we use  the bosonization identity \eqref{eq:bosonization-id} and obtain
\begin{equation}
A(E,x) 
	= 
	\pwsum_{\pm}
	\pwsum_{m_c,m_s>0}
	A_{m_c,m_s}^{(\pm)}(x) \,
	\updelta(E \mp \omega_{m_c,m_s})
	\mathcomma
\end{equation}
with the spectrum of charge and spin excitations
\begin{equation}
\omega_{m_c,m_s} = \frac{E_{\rm gap}}{2} + v_c q_{m_c} + v_s q_{m_s}
\end{equation}
and the spectral weights,
\begin{widetext}
\begin{equation} \label{eq:spect-weights}
A_{m_c,m_s}^{(\pm)}(x)
	= 
		C 
		\left| \sin \left( \frac{\pi}{L} x\right) \right|^{-\sum_{\nu} \alpha_{\nu} \beta_{\nu}}
		\Big(
		2 I^{x,-x}_{m_c,m_s} 
		- \e^{\I 2 k_F^{\pm} x } I^{\pm x, \pm x}_{m_c,m_s}
		- \e^{-\I 2 k_F^{\pm} x } I^{\mp x, \mp x}_{m_c,m_s}
		\Big)
	\mathperiod 
\end{equation}
\end{widetext}
The remaining integrations,
\begin{equation} \label{eq:spect-weights-int}
I_{m_c,m_s}^{x,y}
	=
		\ds\pwprod_{\nu = c,s}
		\pwint_{0}^{T_{\nu}} \! \dfrac{\dd t}{T_{\nu}}
		\e^{\I v_{\nu} q_{m_{\nu}} t}
		\,
		\exp
		\left[
		\pwsum_{k=1}^{m_{\nu}}
		\chi_{\nu,k}(x)\chi_{\nu,k}(y)
		\e^{-\I  v_{\nu} q_{k} t}
		\right]
	\mathcomma
\end{equation}
are performed numerically.
Here, ${T_{\nu} = 2 L / v_{\nu}}$ is the travel time of charge and spin density waves.
$C$ is a cutoff-depend prefactor.


\section{Charge density wave}

\subsection{Charge density wave model}
To calculate the LDOS for a CDW we use standard mean-field theory assuming a local interaction $U$ in a continuum model. We solve the Schr\"odinger equation in a box self-consistently,
\begin{equation} \label{eq:stat_Schrodinger}
\left( -\I v_F \partial_x + U n(x) \right) \psi_n(x)
	 =
		E_n \psi_n(x)
	\,,
\end{equation}
with the Fermi velocity $v_F$ and the electron density,
\begin{equation} \label{eq:mf_density}
n(x)
	 = 
		\pwsum_{n=1}^{N_0} |\psi_n(x)|^2
	\,.
\end{equation}
The LDOS of the interacting electrons is then given by
\begin{equation}\label{eq:ldos_cdw}
A(E,x) 
	= 
	\pwsum_{n \in \mathbb{N}}
	|\psi_n(x)|^2 \,
	\updelta(E - E_n)
	\,.
\end{equation}
The CDW gap opens between the highest occupied $E_{N_0}$ and the lowest empty hole state $E_{N_0+1}$ for attractive interactions ${U < 0}$. 
To produce Fig.~3f we take the value of $v_F$ from the DFT calculation and we fit the interaction constant $U$ to the experimental size of the gap $E_{\rm{gap}}$. Note that a negative $U$ is needed to obtain a CDW. The attractive interaction mimics the effect of optical phonons giving rise to an attractive interaction at momenta $2 k_F$.

\subsection{Room temperature STM measurements}

\begin{figure}[t]
	\centering
		\includegraphics[width=0.475\textwidth]{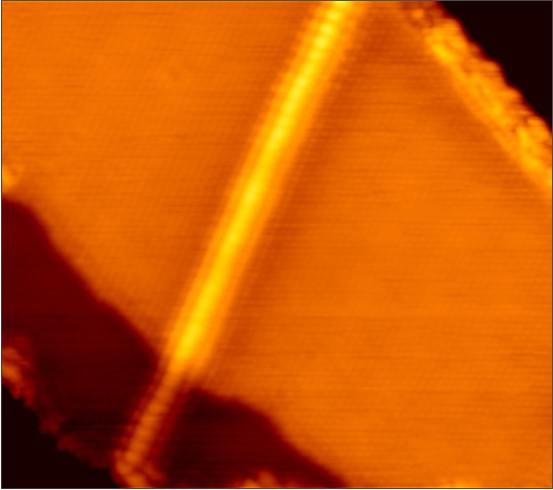}
	\caption{Room temperature STM measurements of a MTB in \mo~crossing a substrate step. 
	STM image of a 4{\textbar}4E MTB ($U=-0.51$~V, $I=0.056$~nA, image size $\left(20\times18\right)~\rm{nm}^2$). The periodic beating pattern of approximately $2a$ is observed best close to the ends of the MTB.
	}
	\label{300K}
\end{figure}

Our room temperature measurements of 4{\textbar}4E MTBs (Fig.~\ref{300K}) reveal that the characteristic beating pattern observed at low temperature persists also at room temperature. Based on the fit of Fig. 3g, a CDW gap could only be of the order of 10 meV. For a Peierls-type CDW this would result in a transition temperature $T_C=33$~K~\cite{Rossnagel2011}, inconsistent with our room temperature observation. Please note that the observation of the beating pattern does not imply the presence of a TLL at room temperature. But it does imply that at least quantization effects resulting from the finite length of the metallic wire persist up to room temperature.

\section{Fourier spectra of the local density of states}

\subsection{Fourier transformation of experimental and theoretical spectra}
We apply a standard Fourier tranformation to our experimental and theoretical spectra and calculate the modulus of the Fourier mode.
This procedure leads to discrete peaks at momenta $2 \pi n/L $.
Note, however, 
that the resulting picture depends on the precise choice of the width~$L$ of the real-space window used for the Fourier transform.
As $L$ determines the discretization in momentum space, this potentially leads to an error of the order of $\pm 2 \pi/L$ in the position and width of the peaks.
We have included only spectra in our Fourier transformation which show a signal clearly associated to the MTB.
For our plots, we use a linear interpolation of the discrete data.

\subsection{Comparison of Fourier spectra of different TLL parameters $K_c$ and of a charge density wave}
\label{varyK}

\begin{figure*}[t!]
	\centering
		\includegraphics[width=0.7\textwidth]{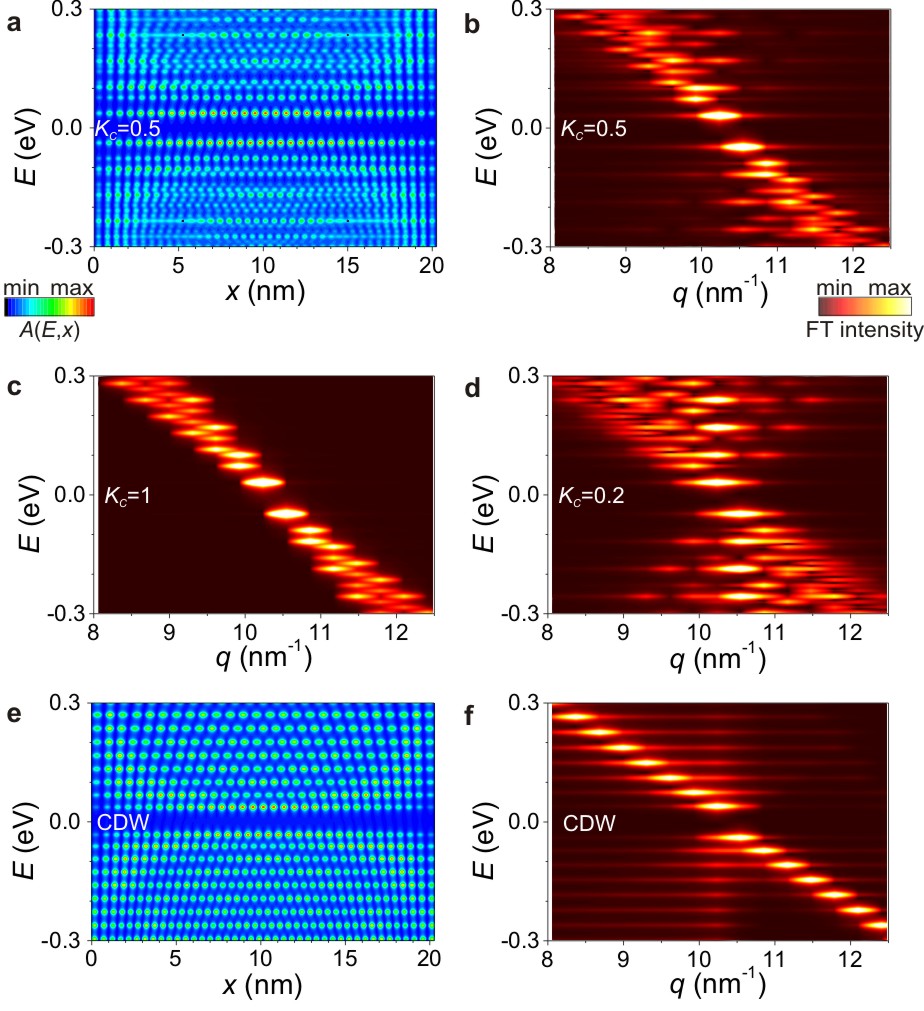}
	\caption{Comparison between various interaction parameters $K_C$ and CDW scenario. \textbf{a} TLL simulation using $K_C=0.5$ as discussed in the main text (${v_c = 0.45~\mathrm{nm \cdot eV}}$, ${v_s = 0.27~\mathrm{nm\cdot eV}}$, $E_{\rm{gap}}=0.08$~eV). \textbf{b} Corresponding FT. \textbf{c} FT of a TLL simulation with $K_C=1$ (nominally non-interacting). \textbf{d} FT of a TLL simulation with $K_C=0.2$ (strong interaction). \textbf{e} CDW simulation using a potential that reproduces the gap $E_{\rm{gap}}= 0.08$~eV (${v_F= 0.25~\mathrm{nm \cdot eV}}$). \textbf{f} Corresponding FT.}
	\label{fig:S4}
\end{figure*}

In the generic interacting case, ${K_c < 1}$, Fourier modes at $2({k_F^{+} \pm q_n)}$ and $2({k_F^{-} \pm q_n)}$ with ${0 \leq n  \leq m_{c}}$ contribute
at a given energy level labeled by the quantum number ${m_{c}}$.
From this selection rule alone, one could expect a symmetric distribution of intensity centered around ${2 k_F^{\pm}}$.
The weights of the Fourier modes, however, are determined by the TLL parameter $K_c$.
We discuss their dependence on $K_c$ for a wire containing the same number of holes as in Fig.~4 (main text).
In Fig.~\ref{fig:S4}a we show the real space image of the LDOS for ${K_c=0.5}$.
Its Fourier transform Fig.~\ref{fig:S4}b reveals which Fourier modes are activated by $K_c$, and to which extent.

In order to make the effect of ${K_c < 1}$ more transparent,
we also show the Fourier spectra for the nominally non-interacting case, ${K_c=1}$, (see Fig.~\ref{fig:S4}c) and for strong interactions, ${K_c=0.2}$ (see Fig.~\ref{fig:S4}d).
Note that we use a fixed ratio ${v_s/v_c= 0.6}$  and a fixed value of ${E_{\rm{gap}} = 0.08\,\mathrm{eV}}$ in all plots as our main goal is to demonstrate the role of matrix elements controlled by $K_c$. In reality a non-interacting system is characterized by ${v_s/v_c=1}$ and the ratio generically shrinks when interactions get stronger. 
For ${K_c=1}$,
only Fourier modes ${2[k_F^{\pm}\pm (q_{m_{c}} + q_{m_{s}}) ]}$ 
with quantum numbers $(m_c,m_s)$
have non-zero weight.
As a consequence the Fourier spectrum displays the linear dispersion
of charge and spin excitations,
${\mp [\frac{E_{\rm gap}}{2} +  v_c (k_F^{\pm} \pm q_{m_c}) +  v_s (k_F^{\pm} \pm q_{m_s}) ]}$.

The most pronounced peaks in Fig.~\ref{fig:S4}c correspond to ``pure'' excitations, e.g., ${(m_c=1,m_s=0)}$.
The remaining intensity is distributed among ``mixed'' excitations, e.g., ${(m_c=1,m_s=1)}$.
For strong interactions (${K_c=0.2}$), a larger number of Fourier modes is activated and the charge peaks are transformed to a more symmetric distribution (as expected from the selection rule).
In this scenario, the most pronounced peaks are found at ${2 k_F^{\pm}}$.
(The number of pure spin peaks does not change since we set ${K_s =1}$.)
Comparison between experimental and theoretical Fourier spectra
allows us to roughly estimate $K_c$ from the number of activated Fourier modes and from the presence or absence of symmetry in the distribution.

In Fig.~\ref{fig:S4}e we show the LDOS of the wire in a CDW state (with the same number of holes and the same value of $E_{\rm gap}$ as used for the TLL plots Fig.~\ref{fig:S4}a--d).
There are three clear differences when compared to the TLL Fourier spectra Fig.~\ref{fig:S4}b--d:
(i) The main peaks in its Fourier spectrum Fig.~\ref{fig:S4}f form a single dispersing band since there is no spin-charge separation.
(ii) The side peaks are only located at ${2 k_F^{-}}$. These indicate the ${2 k_F^{-}}$ scattering processes which lead to the opening of the gap.
(iii) The band slightly curves in the vicinity of the gap while in TLL the bands are strictly linear.


%

\end{document}